\documentclass[a4paper]{jpconf}

\usepackage{graphicx,amsmath,amssymb}
\usepackage{url}

\usepackage[amssymb, cdot]{SIunits}
\usepackage{subfigure}

\begin{document}

\title{Azimuthal Asymmetries From Jets Quenched In Fluctuating Backgrounds}

\author{R Rodriguez$^1$, R J Fries$^2$}

\address{$^1$ Department of Mathematics and Physics, Ave Maria University, Ave Maria FL 34142, USA}
\address{$^2$Cyclotron Institute and Department of Physics and Astronomy, 
   Texas A$\&$M University, College Station, TX 77843, USA}

\ead{ricardo.rodriguez@avemaria.edu, rjfries@comp.tamu.edu}

\begin{abstract}
High momentum jets and hadrons are important probes for the quark gluon
plasma (QGP) formed in nuclear collisions at high energies. We investigate 
how fluctuations in the background density of the QGP and fluctuations in 
the spatial distribution of the hard process create azimuthal asymmetries
of the high momentum hadron spectrum, described by the Fourier coefficients 
$v_n$, $n > 0$ . We estimate the coefficients up to $v_6$ in a simple energy
loss model tuned to single inclusive hadron suppression.
\end{abstract}

With the study of quark gluon plasma in collisions of relativistic heavy ions moving into an 
increasingly accurate quantitative phase it has been found important to
include fluctuations in the space-time structure of the fireball into
calculations of bulk quantities \cite{Alver:2008zza,Alver:2010gr}. These can 
emerge from fluctuations in the initial energy density $\epsilon(x,y)$ in the plane
transverse to the beam axis and leave signature effects on bulk quantities
like the azimuthal asymmetry coefficients $v_n$. For example fluctuations can
lead to a sizeable triangular flow $v_3$ which would be vanishing in an
averaged fireball due to the overall geometry of the nuclear overlap \cite{Alver:2010gr}. At large transverse momentum $P_T$ fluctuations in the 
position of the hard process can also affect observables accessible in current
heavy ion experiments.

Here we have explored the role of such fluctuations on the
suppression of high-$P_T$ hadrons and their generalized azimuthal asymmetry 
coefficients $v_n$. The $v_n$ are defined as a Fourier decomposition of the 
azimuthal angle dependent spectrum
\begin{equation}
  \frac{dN}{P_T d P_T d\Phi} = \frac{dN}{2\pi P_T dP_T} \left[ 1+ 2 \sum_{n>0}
    v_n(P_T) \cos (n\Phi + \delta_n) \right]  
\end{equation}
where the angle $\Phi$ is measured with respect to the reaction plane defined
below and the $\delta_n$ are phases that encode a misalignment with the reaction plane.
We note that for smooth, non-fluctuating fireballs we expect all odd
coefficients $v_1$, $v_3$ etc.\ at midrapidity to vanish for symmetry
reasons.

However, in any given single event the initial energy density will typically exhibit a
non-vanishing triangular eccentricity $\epsilon_3$ which could in turn lead to a
non-vanishing $v_3$. The event-by-event fluctuations in initial energy density
are driven by fluctuations of the positions of nucleons in the initial nuclei
and in the amount of energy deposited around midrapidity for every
nucleon-nucleon collision \cite{Qin:2010pf}. We also expect that the
triangular eccentricity is not correlated with the reaction plane, i.e.\ 
$\delta_3$ should appear random. Similar arguments can be made for other 
$n>0$ and we expect all odd $v_n$ to acquire non-vanishing values with
realistic fluctuations. While all of this has first been discussed for the 
bulk of the fireball \cite{Qin:2010pf} these statements easily transfer to
hard probes. Fluctuations in the energy density lead to fluctuations in energy
loss. In addition, the position of a hard process which creates a hard probe
is subject to fluctuations. 

Systematic measurements of $v_n$ at large momentum could lead to further 
constraints on the type of energy loss prevalent in QGP, and on the size of
the transport coefficient $\hat q$. It can also give an independent handle on 
the size and granularity of initial state fluctuations. Here we report 
on a quantitative study of high momentum azimuthal coefficients using a simple
energy loss model with realistic fluctuations.


We calculate high momentum hadron spectra using our simulation package PPM 
\cite{Rodriguez:2010di,Fries:2010jd}. It samples initial momentum distributions of quark
and gluon jets from a perturbative calculation and propagates leading partons
through a given background fireball. Different energy loss models can be
employed. Here we will show results from a simple LPM-inspired (sLPM) 
deterministic energy loss model $dE/dx \sim\hat q x$ where $\hat
q$ scales with the 3/4th power of the local energy density
\cite{Rodriguez:2010di}. As a cross check we will sometimes also use the 
non-deterministic Armesto-Salgado-Wiedemann  (ASW) model 
\cite{Salgado:2003gb,Dainese:2004te}. Fitted to the same 
experimental data on single hadron suppression these models cover a wide range
of values for $\hat q$. PPM will eventually fragment leading partons into
hadrons and all results here will be shown for pions.

\begin{figure}[tb]
  \mbox{
        \includegraphics[width=3.9cm,height=3.9cm,angle=0]{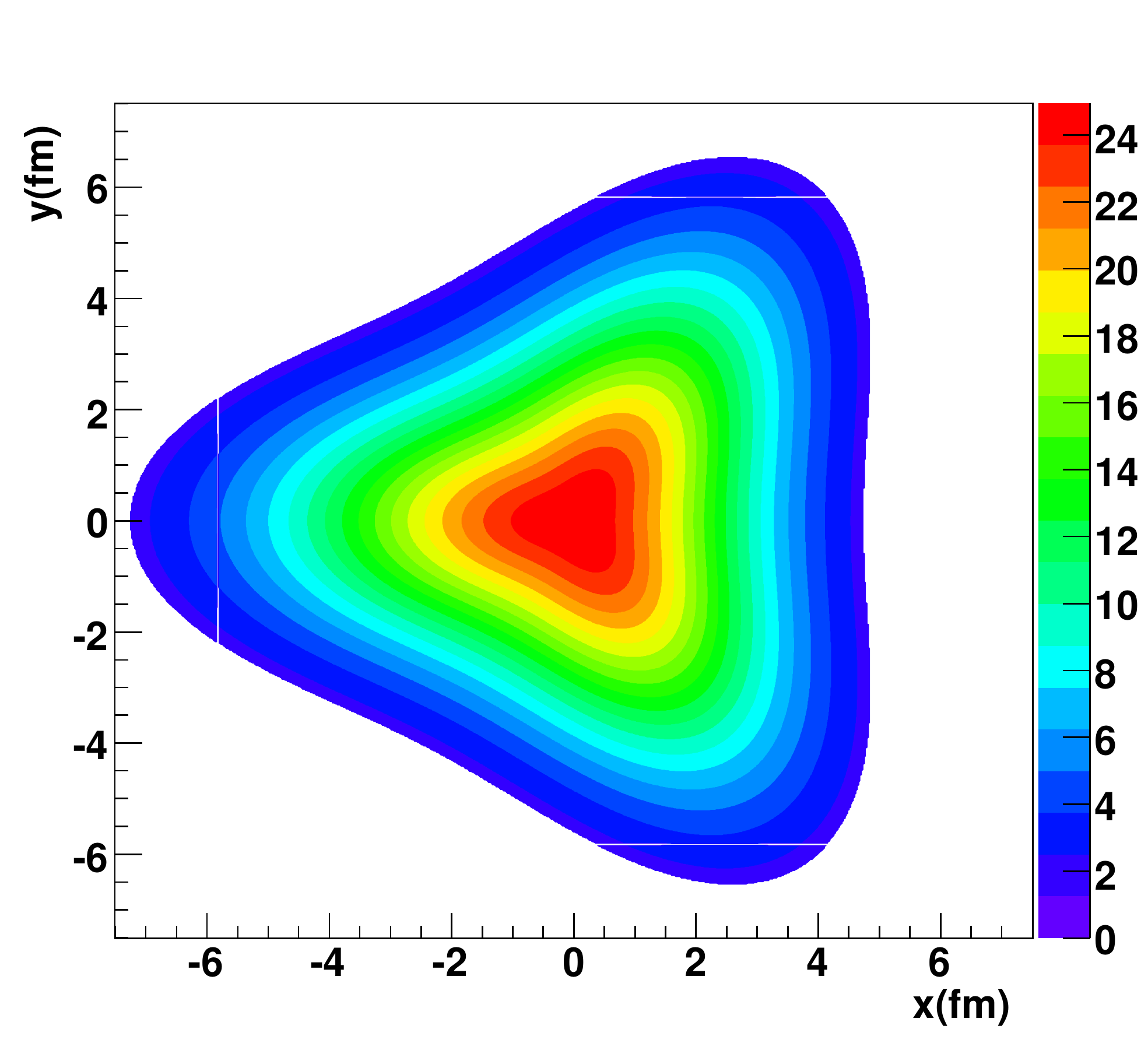}
         \includegraphics[width=3.9cm,height=3.9cm,angle=90]{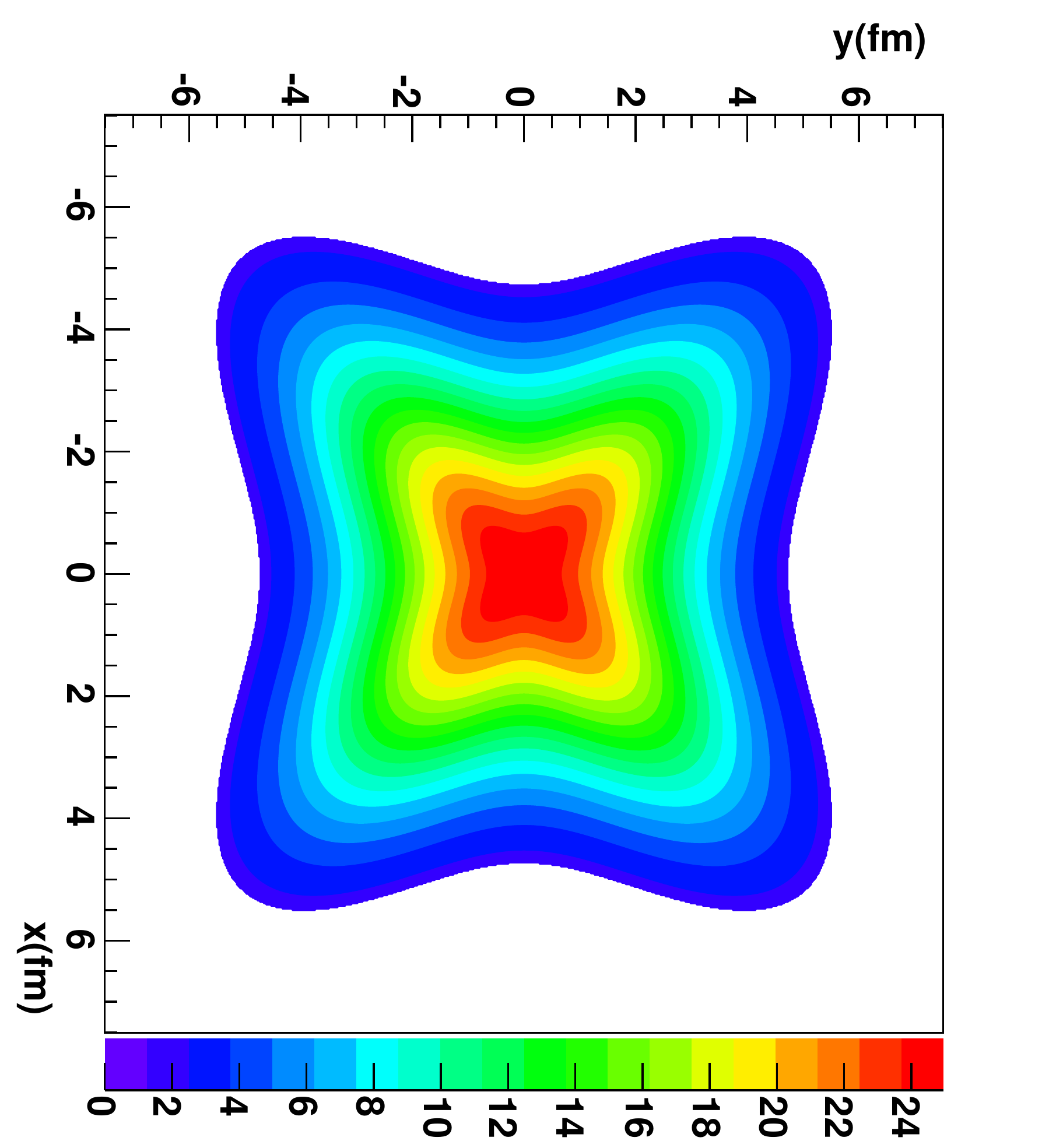}
          \includegraphics[width=3.9cm,height=3.9cm,angle=90]{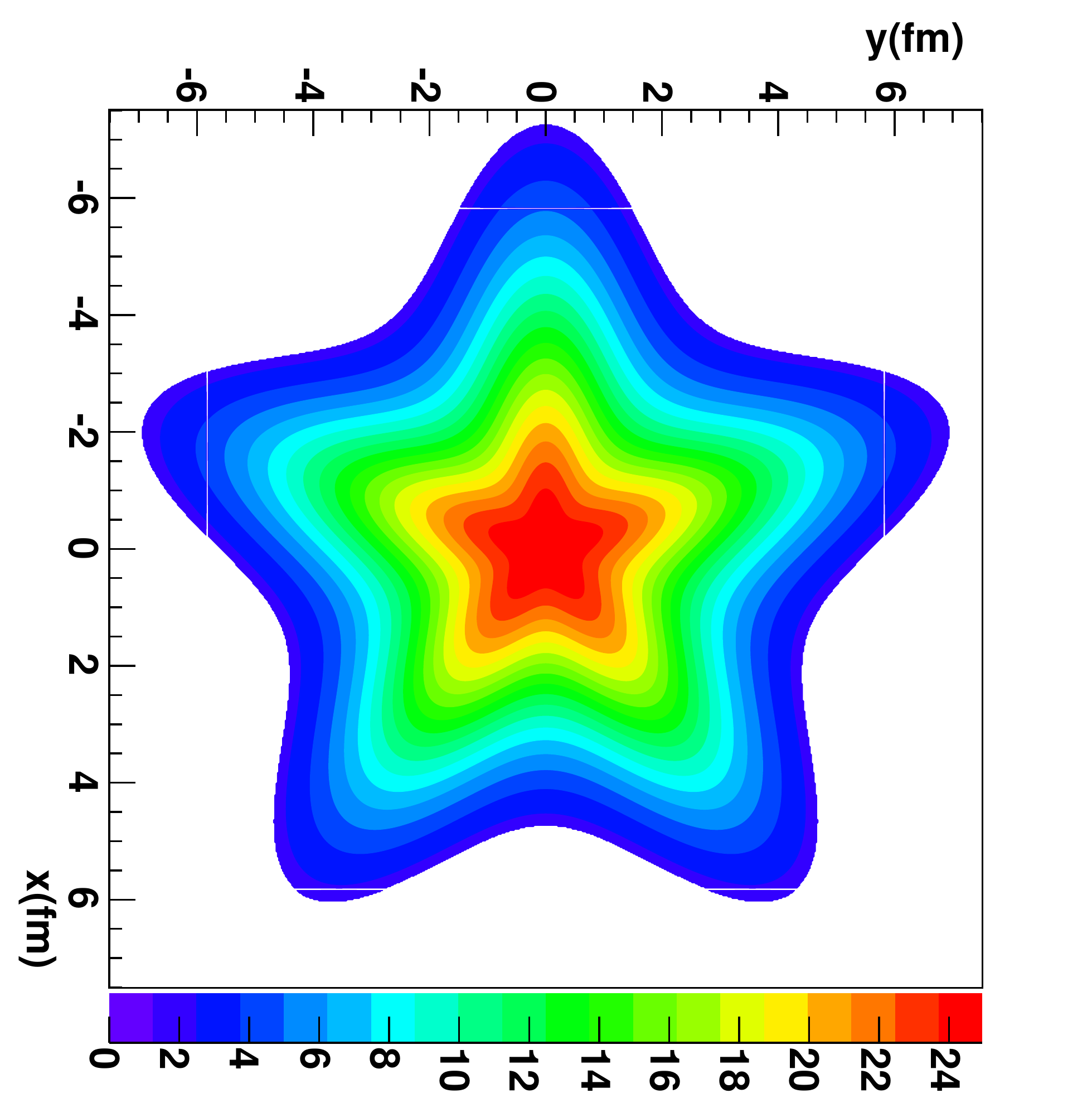}
          \includegraphics[width=3.9cm,height=3.9cm,angle=90]{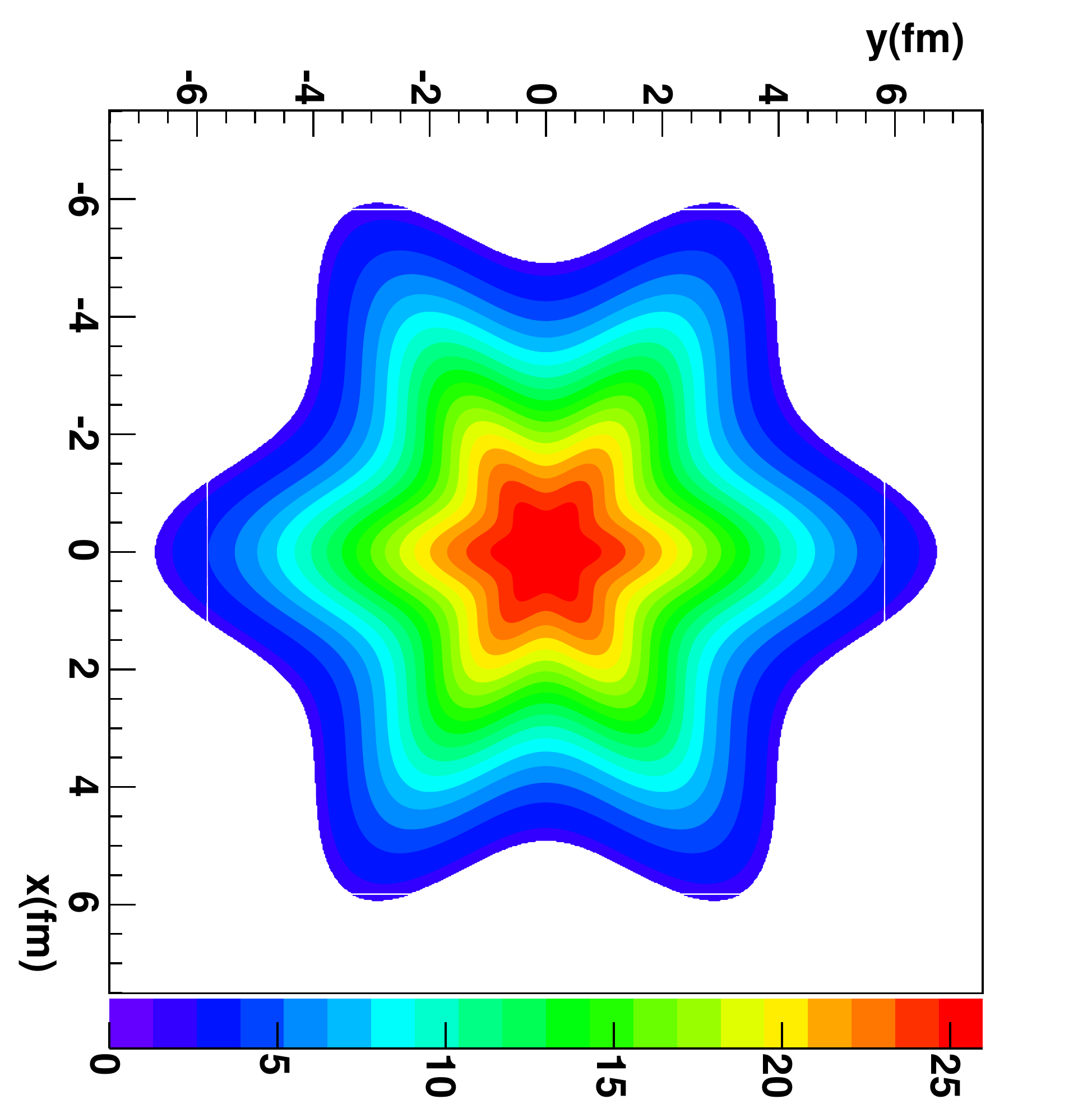}}
  \caption{ Engineered events with $n=3,4,5,6$}
  \label{fig:eng_events}
\end{figure}

We have used the Glauber Monte Carlo generator GLISSANDO \cite{Broniowski:2007nz} 
to produce an ensemble of  Au+Au events at top RHIC energy using the 
three different centralities $b= 3.2,\ 7.4$ and $11\,\femto\meter$. 
However, first we check the general relation between the spatial
eccentricities $\epsilon_n$ \cite{Petersen:2010cw}
\begin{equation}
 \epsilon_n=\frac{\sqrt{\left\langle r^n\cos(n\phi)\right\rangle^2+\left\langle r^n\sin(n\phi)\right\rangle^2}}{\left\langle r^n\right\rangle}\label{eq:ecc_generalization}
\end{equation}
and the azimuthal asymmetry coefficients $v_m$
using ``engineered'' events with particular fixed eccentricities. These are created
with the energy density modeled as simple Gaussians in the transverse plane
with a $\cos n\phi$ undulation of the mean square radius, as shown in the
examples in  Fig.\ref{fig:eng_events}.

\begin{figure}[tb]
  \mbox{
        \includegraphics[width=3.9cm,height=3.9cm,angle=90]{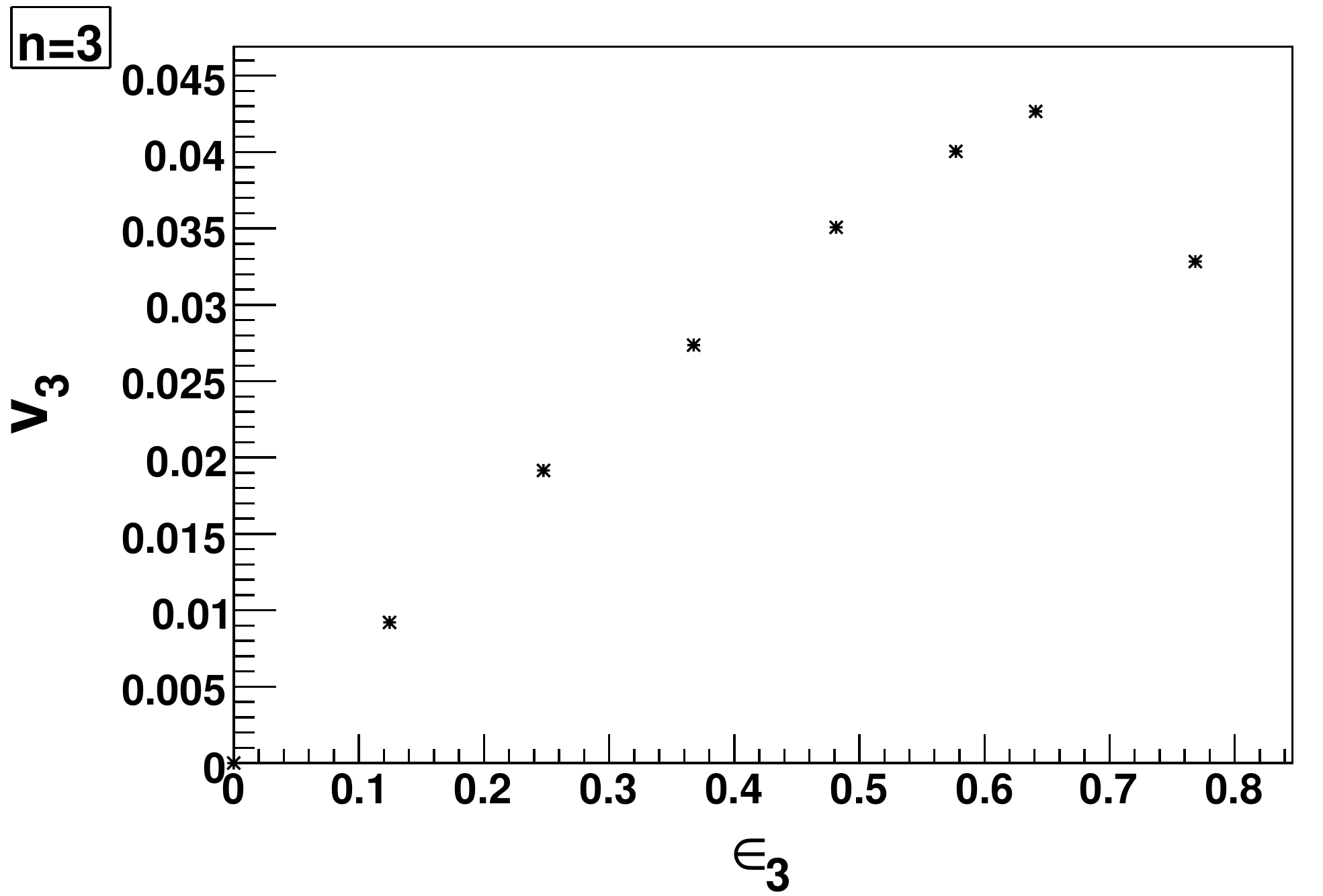}
         \includegraphics[width=3.9cm,height=3.9cm,angle=90]{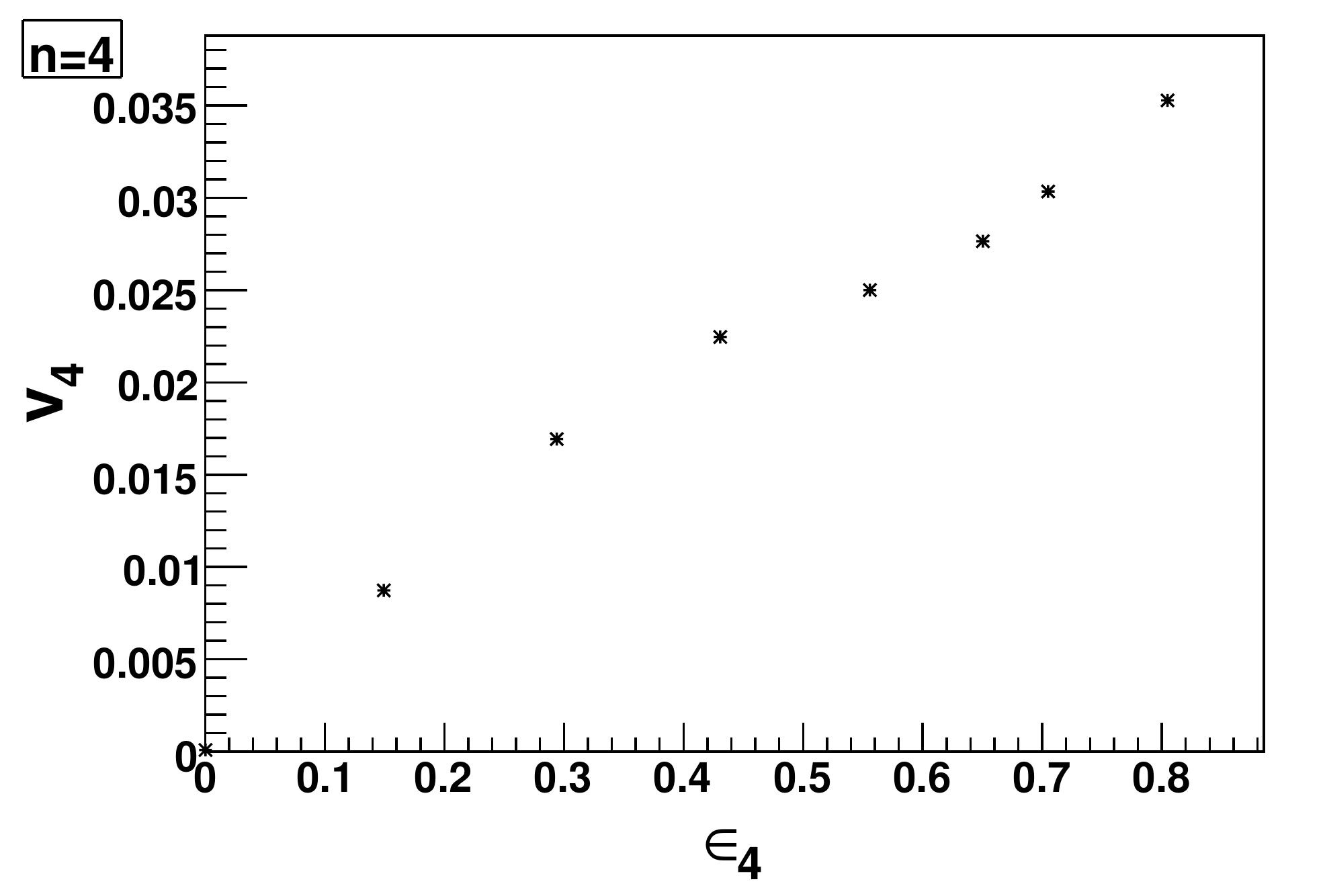}
          \includegraphics[width=3.9cm,height=3.9cm,angle=90]{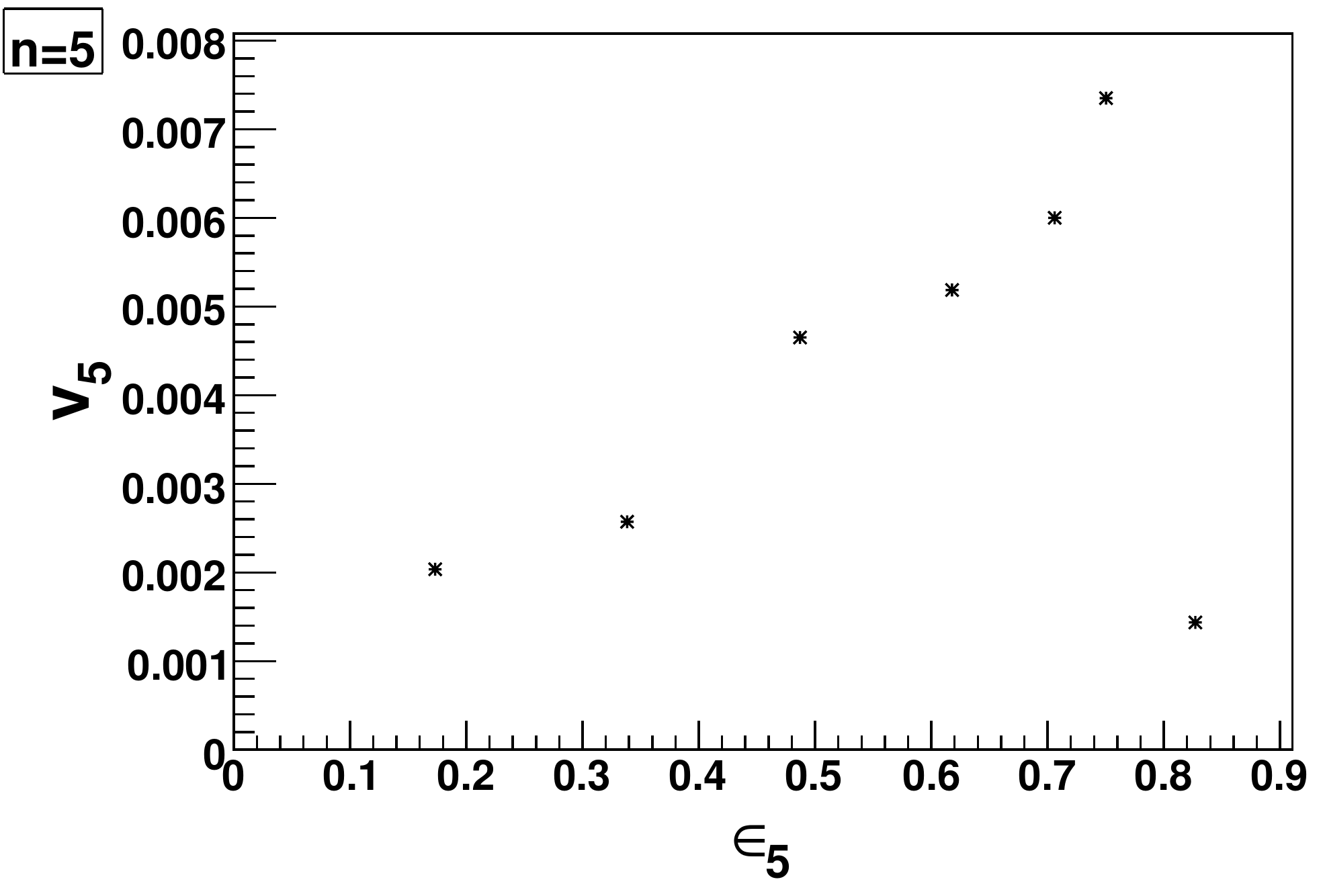}
          \includegraphics[width=3.9cm,height=3.9cm,angle=90]{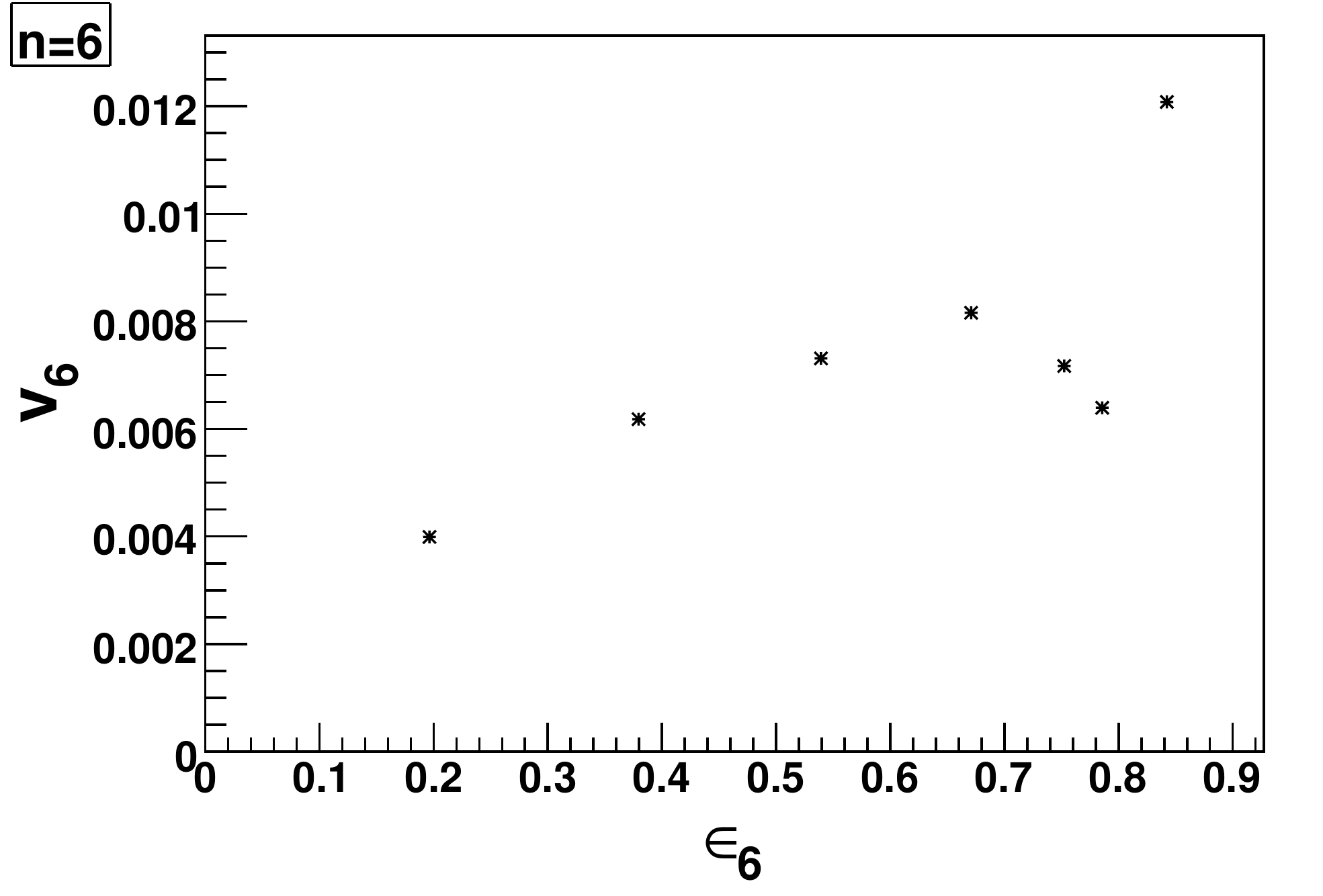}}
  \caption{ Azimuthal asymmetry $v_n$ vs eccentricity $\epsilon_n$ in
    engineered events for $n=3$, 4, 5, 6.}
  \label{fig:vnen_results}
\end{figure}

\begin{figure}[tb]
\begin{center}
 \mbox{
        \includegraphics[width=4.5cm,height=4.5cm,angle=90]{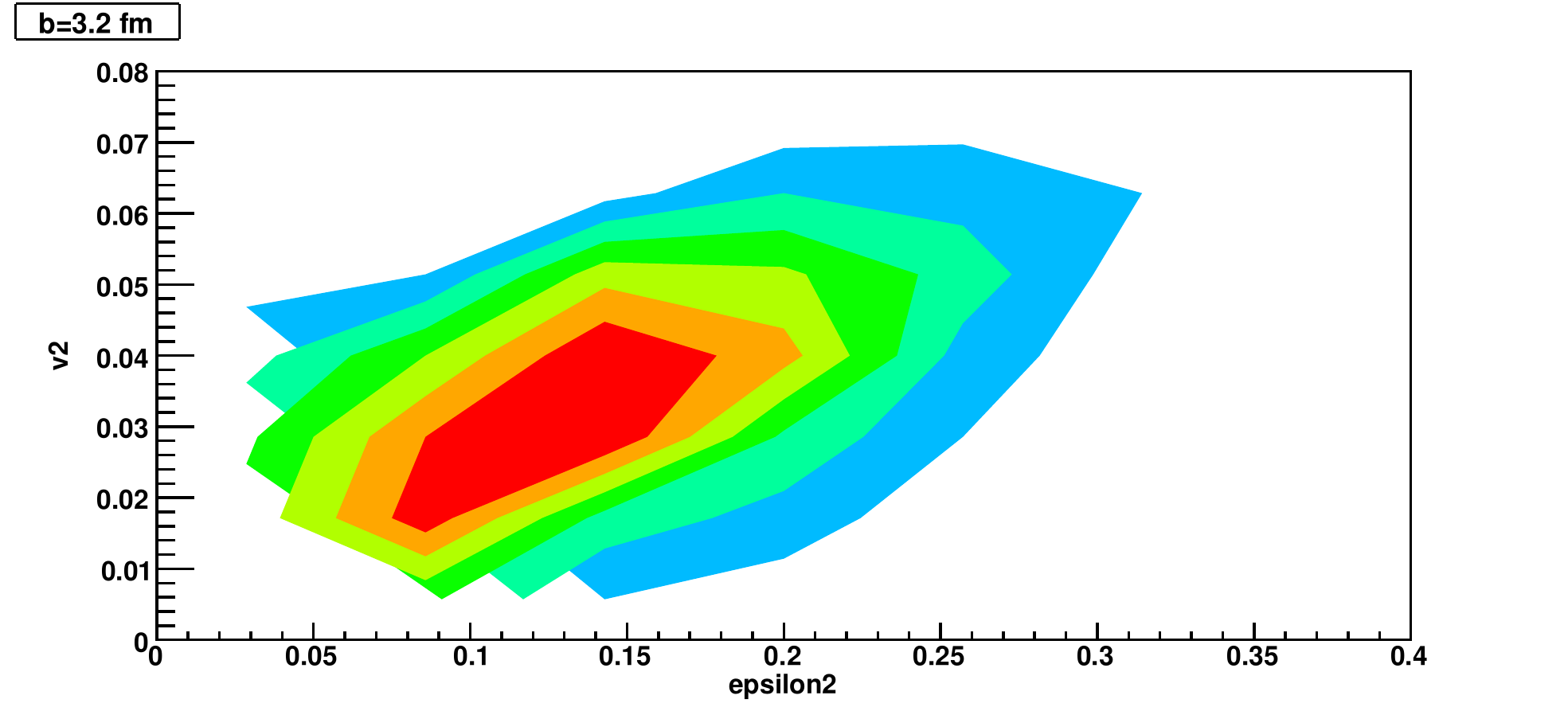}
         \includegraphics[width=4.5cm,height=4.5cm,angle=90]{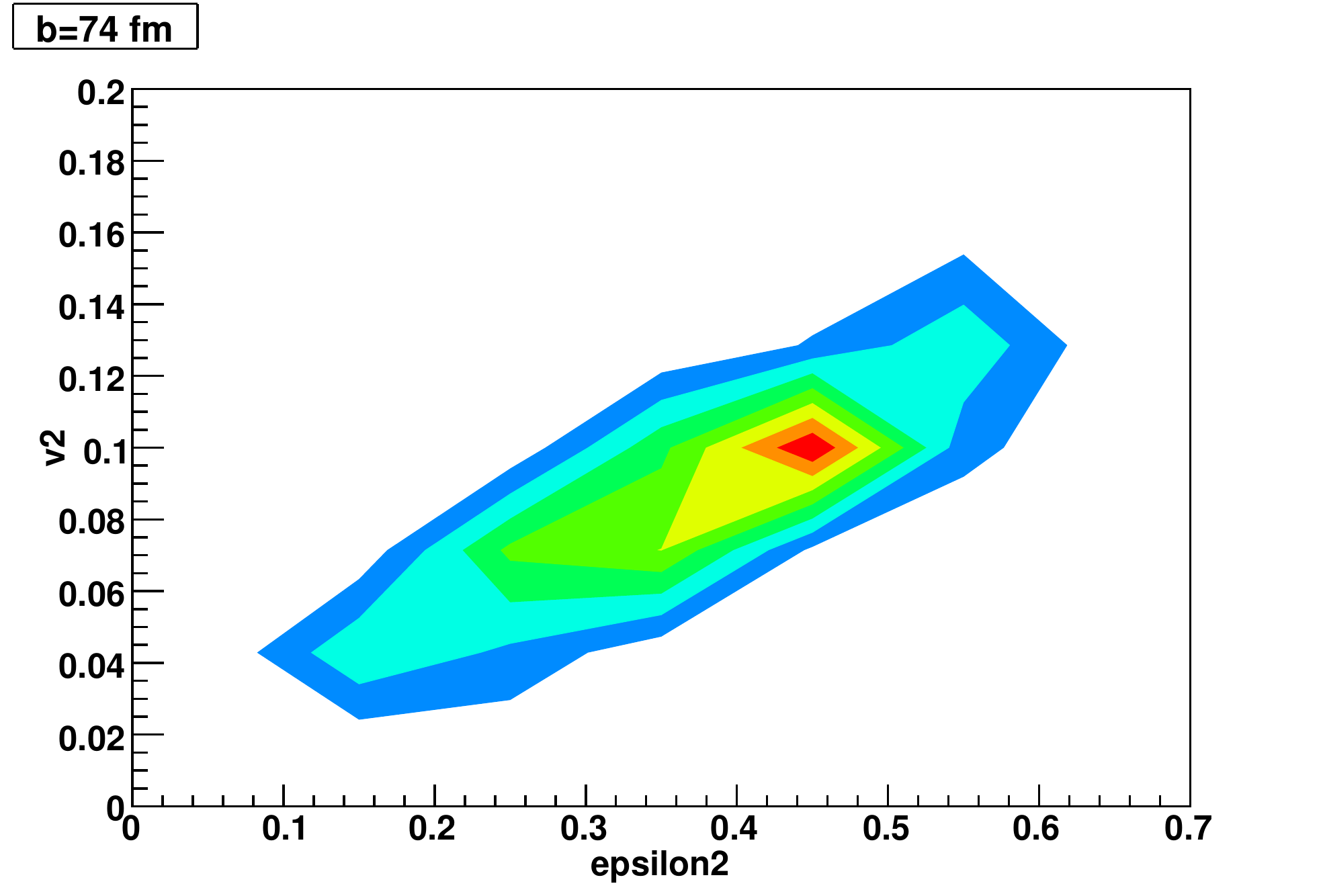}
          \includegraphics[width=4.5cm,height=4.5cm,angle=90]{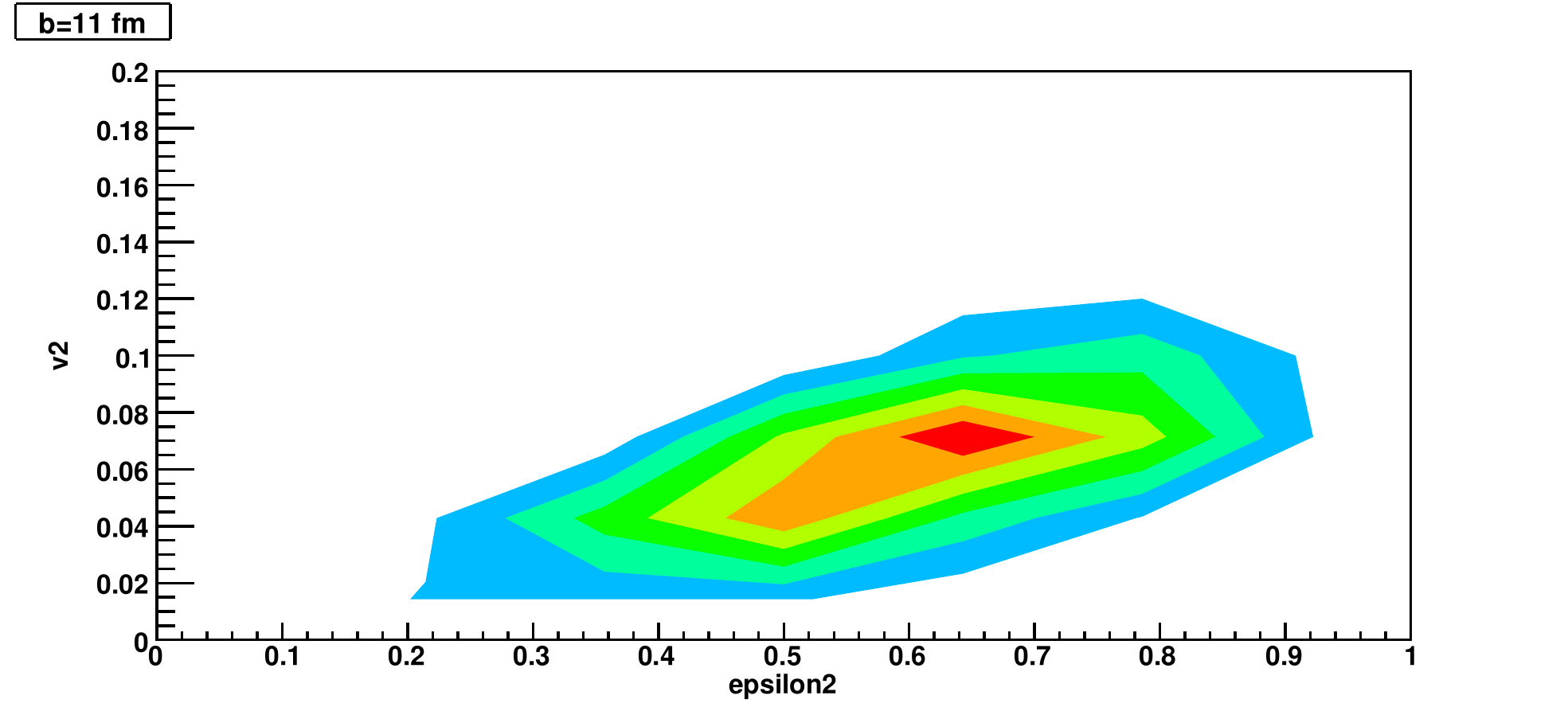}}
  \caption{Correlation between $v_2$ and $\epsilon_2$ for an ensemble of 
  Au+Au collisions at top RHIC energy in GLISSANDO for three impact parameters 
  3.2, 7.4 and 11 fm.}
 \label{fig:vnen_results_gl}
\end{center}
\end{figure}

First we scanned the space $(v_n,\epsilon_m)$, $n,m=1,6$ in search of
correlations. As expected we find non-zero $v_n$ for a given $\epsilon_m$ only if 
$n=im$ where $i>0$ is an integer. Next we explored the scaling of $v_n$ with
the size of the eccentricity $\epsilon_n$. We expected a monotonically
increasing function $v_n(\epsilon_n)$ which can be seen confirmed in Fig.\,\ref{fig:vnen_results}. 
Deviations from monotony only occur for
unrealistically large eccentricities.
These basic results should hold if realistic fluctuations are considered.  
Fig.\,\ref{fig:vnen_results_gl} shows the correlation between $v_2$ and
$\epsilon_2$ for our ensemble of GLISSANDO events for all 3 impact parameters
for Au+Au collisions at top RHIC energies.
The basic linear correlation persists for $n=2$ but is washed out. Correlations for
$n>2$, not shown here, are much more weakened to a point that makes it hard to 
predict $v_n$ for a known $\epsilon_n$.
All results shown are for sLPM energy loss but the corresponding result
for ASW show no noticeable difference, making the conclusions rather robust
against large variations in the microscopic origin of energy loss.

\begin{figure}[tb]
  \mbox{
        \includegraphics[width=3.9cm,height=3.9cm,angle=90]{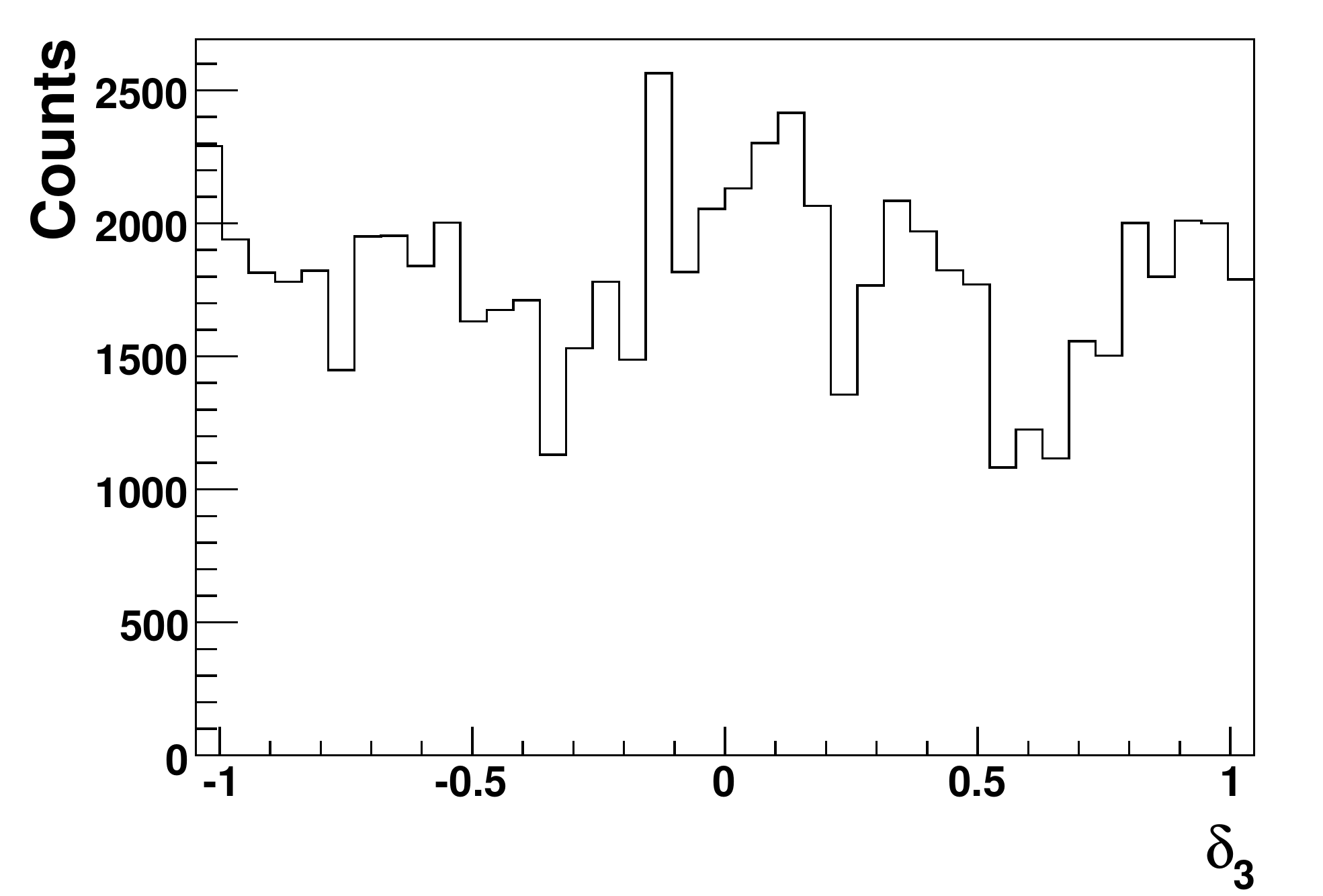}
         \includegraphics[width=3.9cm,height=3.9cm,angle=90]{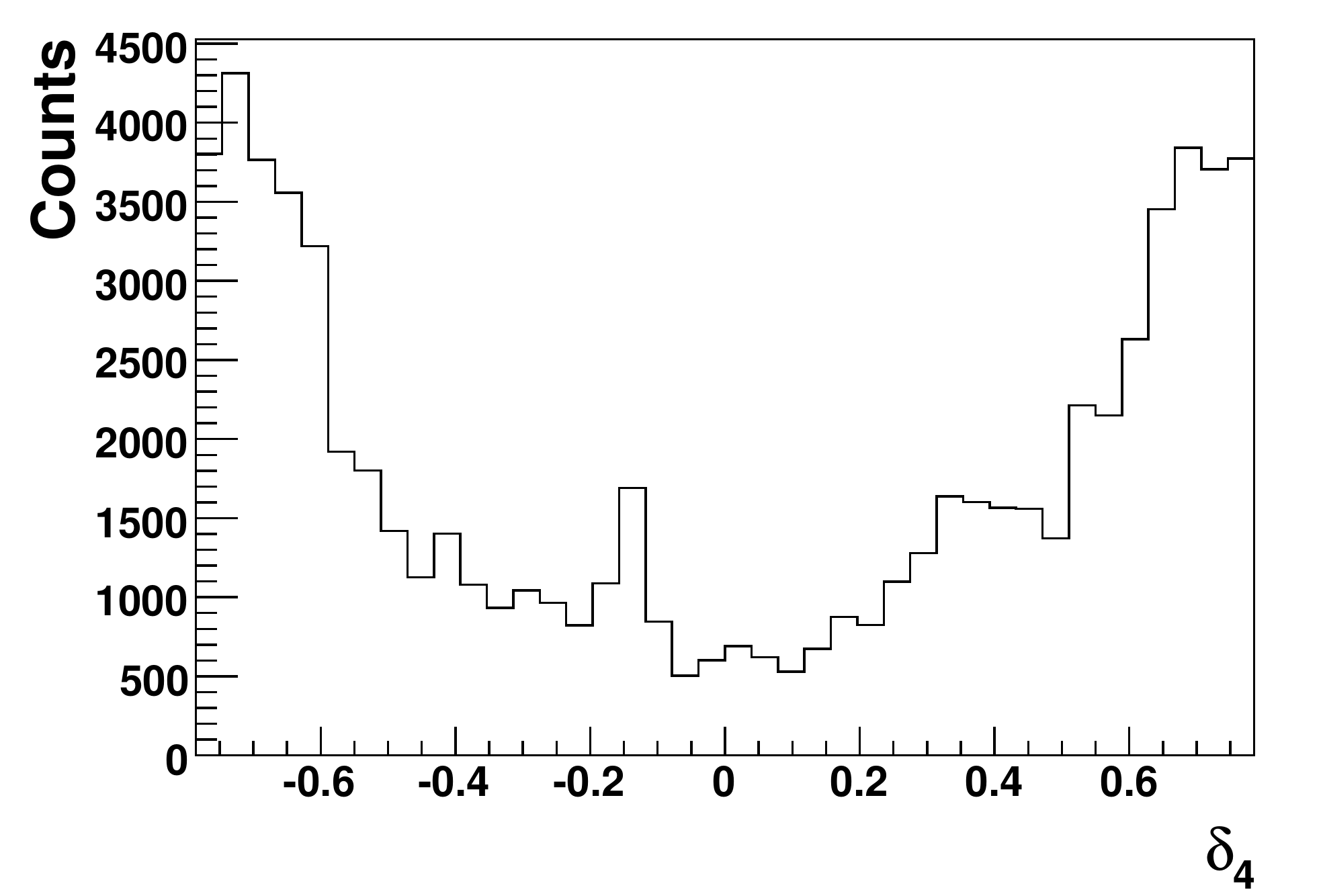}
          \includegraphics[width=3.9cm,height=3.9cm,angle=90]{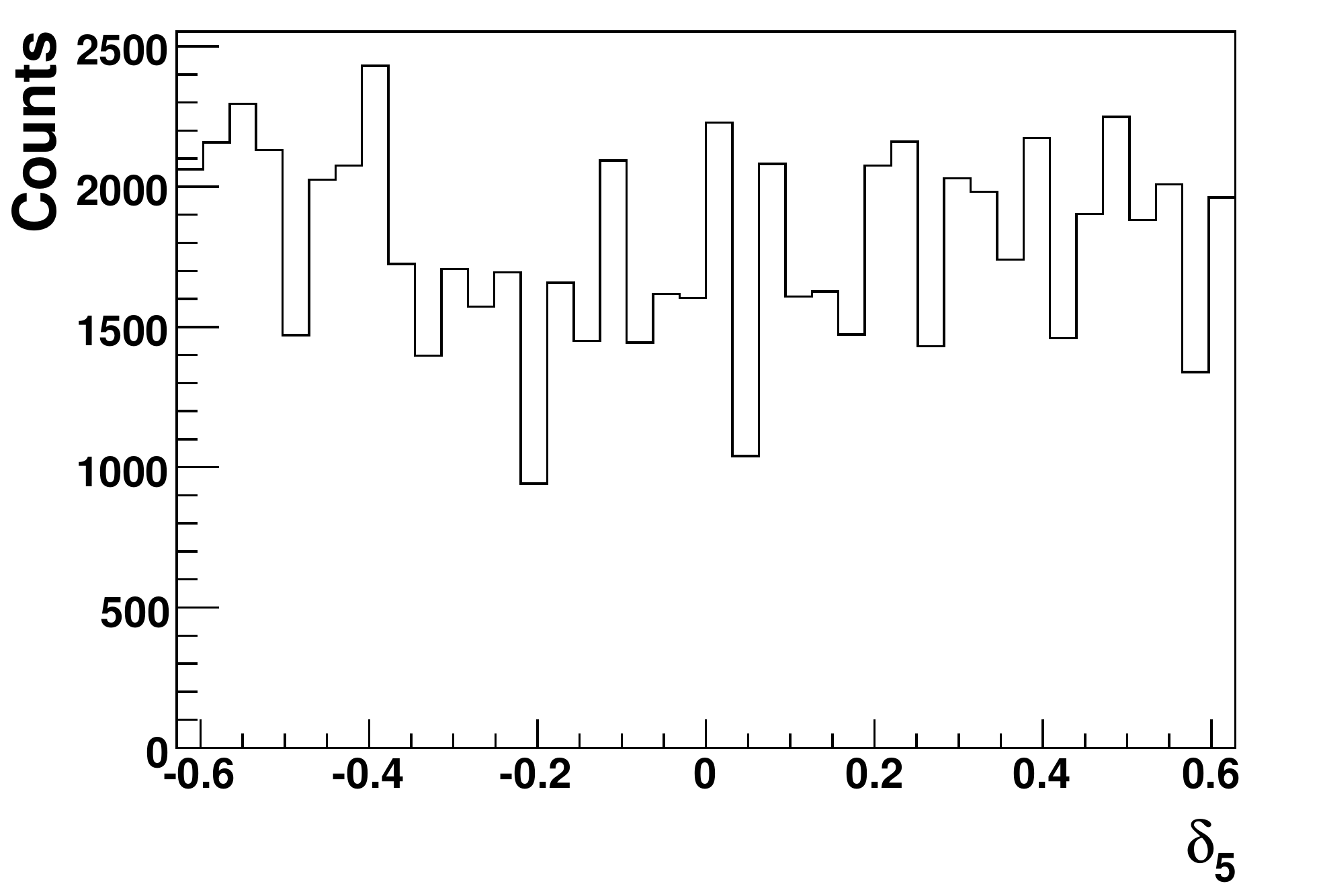}
          \includegraphics[width=3.9cm,height=3.9cm,angle=90]{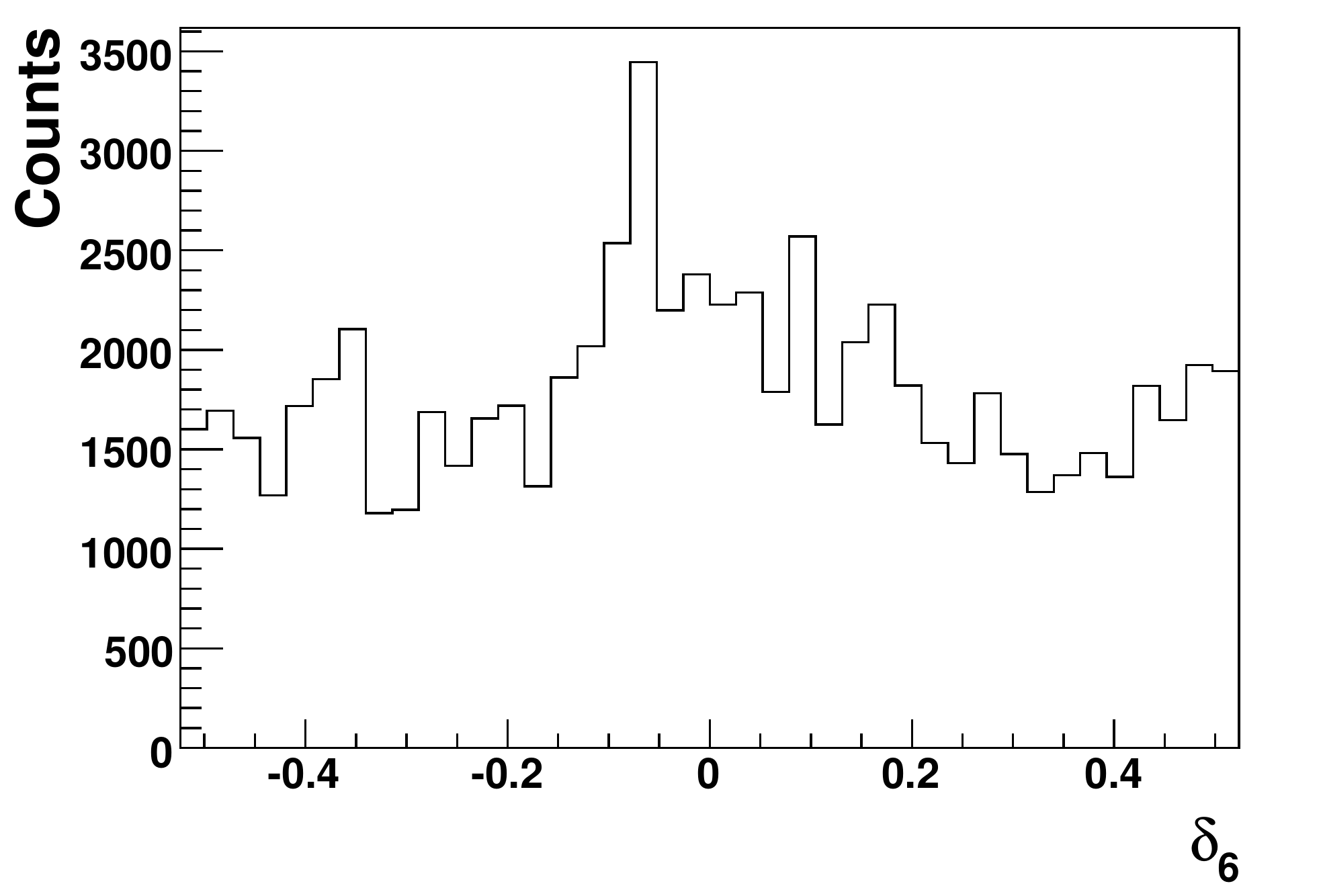}}
  \caption{Distribution of the phases $\delta_n$ for $n=3$, 4, 5, 6 in our ensemble of
    Au+Au events.}
  \label{fig:delta_results}
\end{figure}

We determined the phases $\delta_n$ for high momentum pions in our ensemble of 
GLISSANDO events for three different impact parameters. In 
Fig.\,\ref{fig:delta_results} we plot those phases relative to the reaction
plane determined by the eccentricity $\epsilon_2$.  In other words our
definition of a reaction plane is given by the fundamental initial
ellipticity, which in general deviates from the plane defined by beam axis and
impact vector. We observe that $\delta_3$ and $\delta_5$  are randomly
distributed, so there is no correlation between the reaction plane and the 
fluctuations which create $\epsilon_3$ or $\epsilon_5$. This had also been
found for the bulk azimuthal asymmetries before \cite{Petersen:2010cw}. 
$\delta_4$ and $\delta_6$ on the other hand show a correlation with the
reaction plane which is consistent with $v_4$ and $v_6$ receiving
contributions from $\epsilon_2$.

\begin{figure}[tb]
  \mbox{
        \subfigure[{sLPM, $b=3.2\,\femto\meter$}]{\includegraphics[width=3.5cm,height=3.9cm,angle=90]{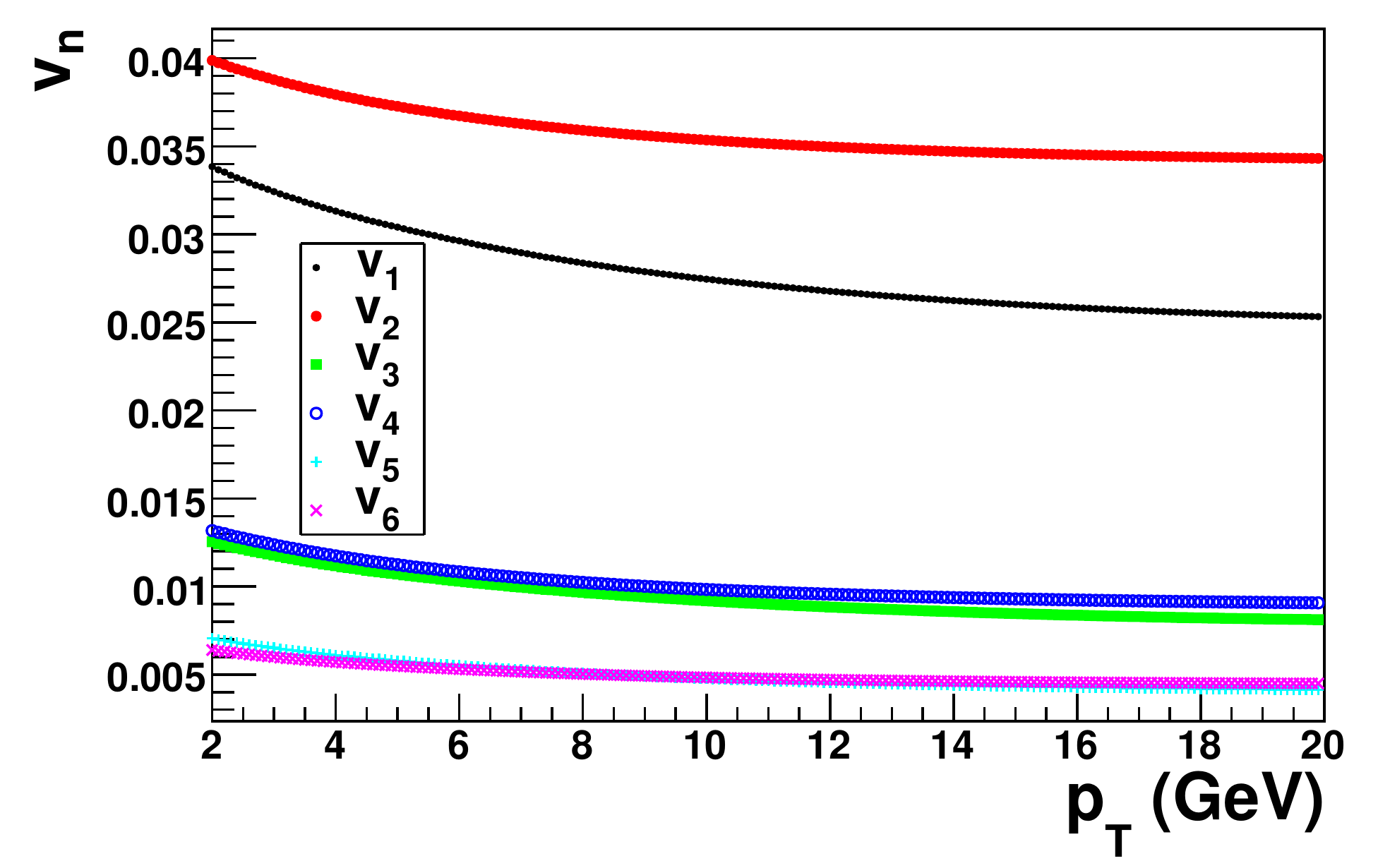}}
        \subfigure[{ASW, $b=3.2\,\femto\meter$}]{\includegraphics[width=3.5cm,height=3.9cm,angle=90]{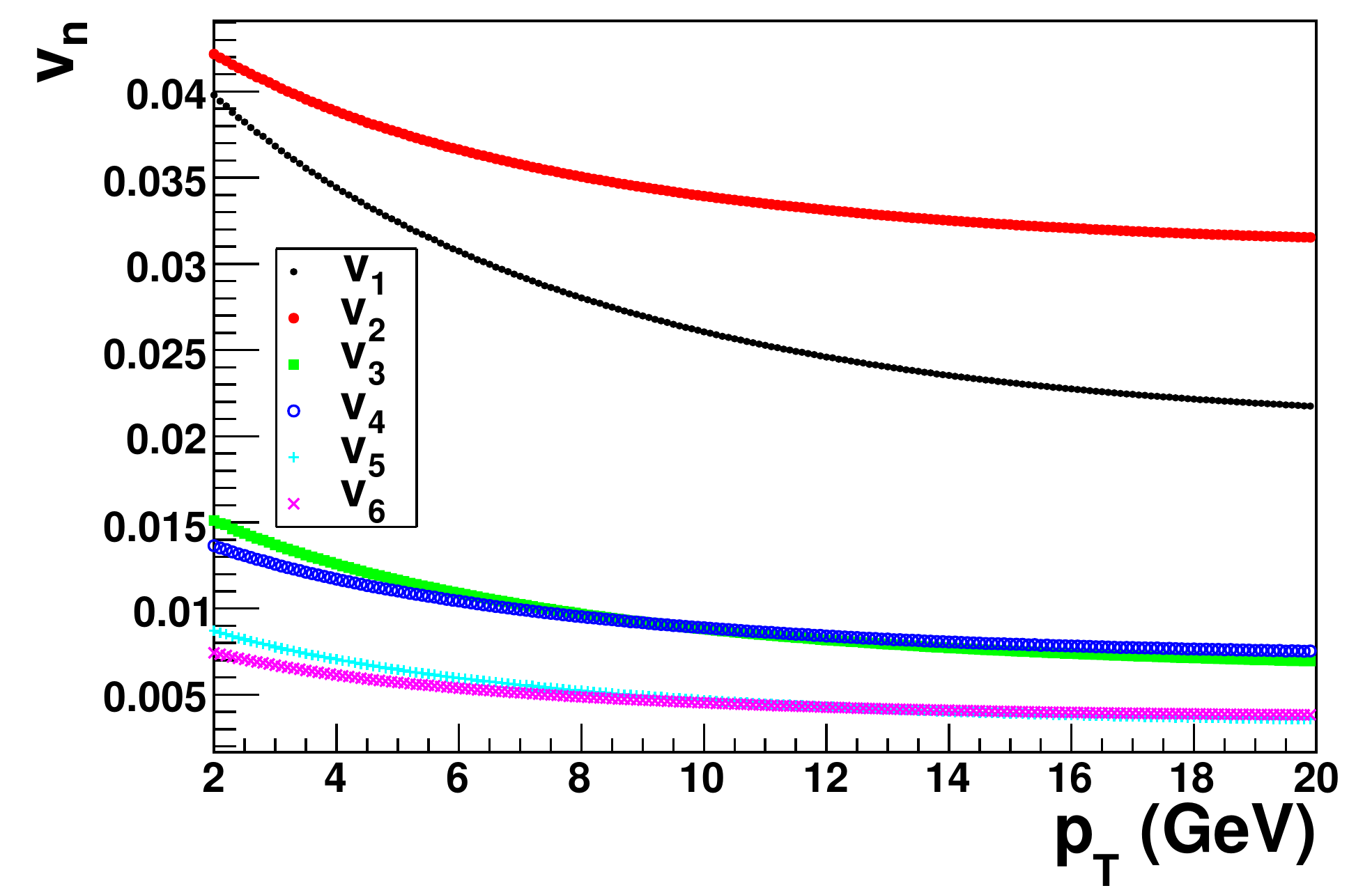}}
        \subfigure[{sLPM, $b=7.4\,\femto\meter$}]{\includegraphics[width=3.5cm,height=3.9cm,angle=90]{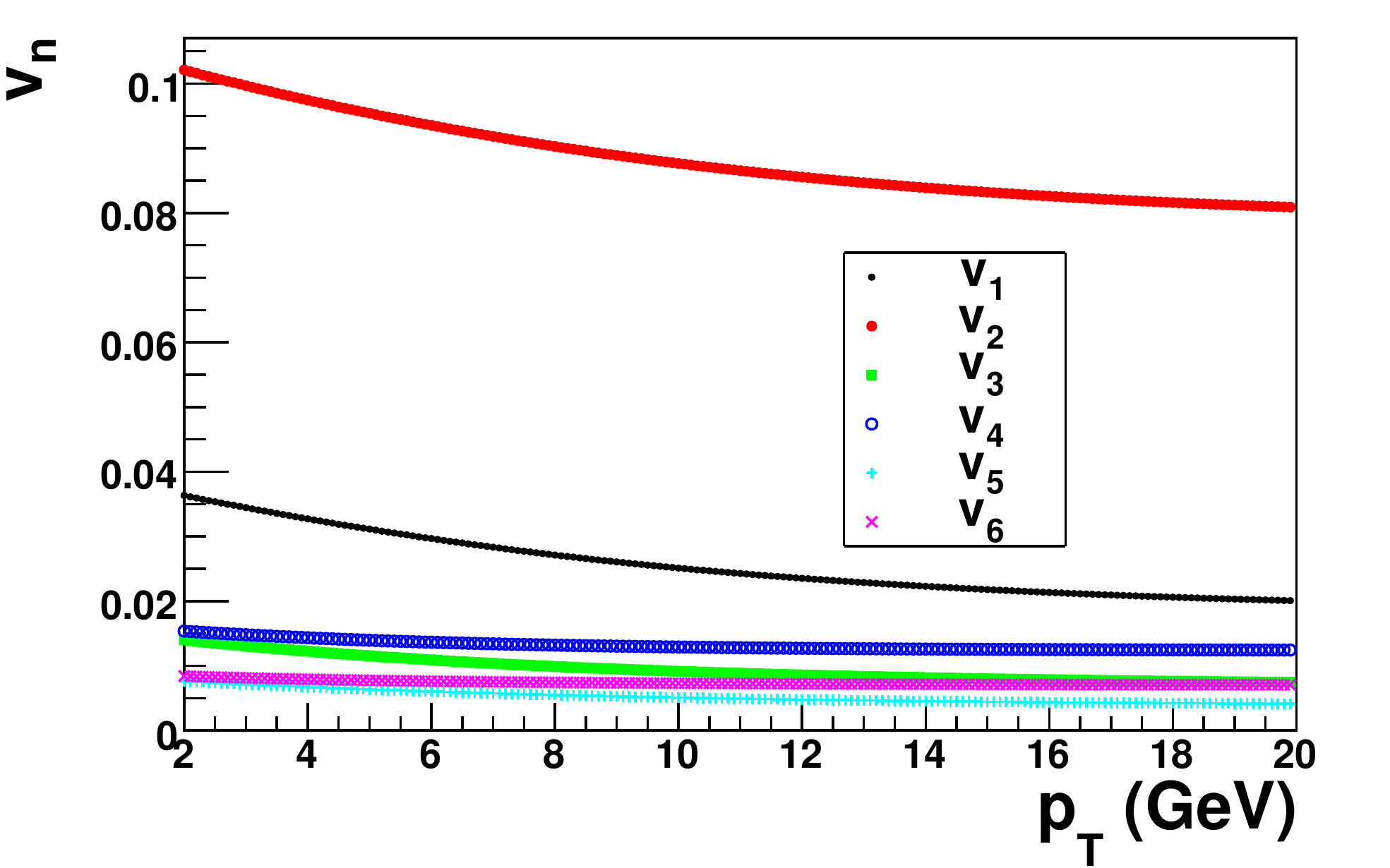}}
        \subfigure[{ASW, $b=7.4\,\femto\meter$}]{\includegraphics[width=3.5cm,height=3.9cm,angle=90]{vnvspt_slpm_74}}}
  \caption{Pion $v_n$ vs $P_T$ for two impact parameters for Au+Au collisions at
    top RHIC energy, calculated with either sLPM or ASW energy loss.}
  \label{fig:vnpt_results}
\end{figure}

The transverse momentum dependence of the coefficients $v_n$ for pions for
our ensemble of GLISSANDO Au+Au events is shown in Fig. \ref{fig:vnpt_results}.
We observe two hierarchies of coefficients, one being $v_2>v_4>v_6$ and the second
one being  $v_1>v_3>v_5$. $v_2$ is always the largest coefficient, even in the
most central events. Interestingly $v_1$ is non-zero and the second largest
coefficient, beating $v_3$ and $v_4$ by more than a factor 2. Momentum
conservation dictates a sum rule for $v_1$ integrated over $P_T$. The recoil of the
medium in which energy is lost would lead to a negative $v_1$ at
lower momentum. Such a back reaction is not included in this
calculation. $v_1$ at intermediate and large $P_T$ could be sensitive to
the mechanism of medium recoil. Generally we point out that our results start
to become unreliable below 4 to 6 \giga\electronvolt\, since hydrodynamic expansion was not included in the calculation.
Fig. \ref{fig:vnpt_results} shows the results for both sLPM and ASW energy
loss. We have to conclude that azimuthal asymmetry coefficients are not
particularly useful to discriminate between energy loss models.
We also find that the $P_T$ dependence of the coefficients becomes
rather weak at large momenta.

To summarize, in this study we have explored higher order azimuthal asymmetry 
coefficients at large momentum in heavy ion collisions. We find that in general $v_n$ rises
with $\epsilon_n$ but this correlation weakens for larger $n$. We also find
that there are only a few cross correlations between $v_n$ and $\epsilon_m$ for $n\ne m$. 
We have also classified the preferred angular orientation of the azimuthal 
asymmetries as given by the phases $\delta_n$ with respect to the reaction
plane. We find a decorrelation with the reaction plane for all odd $n$.
Finally we have made predictions for $v_n$ as a function of $P_T$ in two
different energy loss models. We find mostly consistent results between those
two models with $v_2$ being the largest coefficient followed by $v_1$.
In general the $v_n$ carry geometrical information and measurements
at large momentum can be complementary to those for bulk observables, but they
seem less useful to distinguish different energy loss models.

This work was supported by NSF CAREER Award PHY-0847538 and by the JET Collaboration and DOE grant DE-FG02-10ER41682.

\end{document}